# Fault Detection in IEEE 14-Bus Power System with DG Penetration Using Wavelet Transform

[1]Prakash K. Ray, [2]B. K. Panigrahi, [2]P. K. Rout

[1]*Dept. of Electrical and Electronics Engineering, IIIT, Bhubaneswar, India*

[2] *Dept. of Electrical Engineering, SOA University, Bhubaneswar, India*

[3]Asit Mohanty, [4]Harishchandra Dubey

[3] *Dept. of Electrical Engineering, CET, Bhubaneswar, India*

[4] *Dept. of ECE, The University of Texas at Dallas, USA*

ABSTRACT: Wavelet transform is proposed in this paper for detection of islanding and fault disturbances distributed generation (DG) based power system. An IEEE 14-bus system with DG penetration is considered for the detection of disturbances under different operating conditions. The power system is a hybrid combination of photovoltaic, and wind energy system connected to different buses with different level of penetration. The voltage signal is retrieved at the point of common coupling (PCC) and processed through wavelet transform to detect the disturbances. Further, energy and standard deviation (STD) as performance indices are evaluated and compared with a suitable threshold in order to analyze a disturbance condition. Again, a comparative analysis between the existing and proposed detection is studied to prove the better performance of wavelet transform.

## 1 INTRODUCTION

Distributed generations (DGs) being regarded as small-scale power resources, which are basically installed near to the loads or can be connected to grid if required. These sources are gaining a great popularity now-a-days because of deregulation and restructuring of the modern power system (Onara et al. 2008, Kim et al. 2008). Of course, the penetration levels of these DGs vary depending upon the location and availability of natural resources. But, with the increase in penetration level, some issues like islanding and fault disturbances become more vital and complex to detect because of uncertain characteristics of renewable resources like solar and wind energy. Islanding is a phenomenon usually occurring when the utility grid is being isolated from the DG because of some abnormal conditions, but during the same condition the DG continues to feed power to loads connected. Similarly fault may occur at any bus or near any bus of the system because of symmetrical and unsymmetrical faults like L-G, L-L, L-L-G, L-L-L, L-L-L-G etc. So once occur, they have to be detected as quickly as possible and corrective measures are to be taken to protect the loads as well as the power system (Ray et al. 2010, Fernandez et al. 2002).

In the literature, many methods were suggested in order to detect the disturbances (Yadav et al. 2014, Jang et al. 2004). The design and influence of multiple staged inverters in detecting the islanding events are presented in ( Yadav et al. 2014). The effect of interfacing control and non-detection zones for islanding detection is discussed in (Zeineldin et al. 2006). Voltage unbalance (VU) and total harmonic distortion (THD) are also taken as indices to detect islanding (Jang et al. 2004). But, if the threshold is not selected properly, the detection becomes difficult sometimes under some operating conditions.

Artificial intelligence techniques like artificial neural network (ANN), fuzzy, adaptive neuro-fuzzy inference system (ANFIS), and support vector machines (SVM) are being used by the researchers for detection and classification of fault disturbances.

In this context, different signal processing techniques are being implemented out of them wavelet transform (WT) is an efficient tool for detection based on time-frequency localization (Dubey et al. 2011, Cheng et al. 2008). WT is very versatile in identifying disturbance features which can easily detect any irregularities in the signal. Comparative detection ability between the existing methods like Fourier Transform (FT), Short time Fourier Transform (STFT) and wavelet transform is carried out with the help of performance indices; standard deviation and energy to prove the better performance of the proposed transform. The work is described section wise as follows; modeling of the IEEE 14-bus system is described in Section 2, WT as signal processing based detection tool is presented in Section 3. Then, the MATLAB based simulation results with descriptions are given in Section 4. Finally, concluding remarks are presented in Section 5.

## 2 IEEE 14-BUS POWER SYSTEM

The penetration level of different renewable power is increasing with time, thereby; the hybrid power system becomes more complex. Hence, the design, operation and control of power system are becoming more challenging. Suitable methods are to be adopted to identify the normal as well as abnormal operating conditions. While operating in grid-connected mode, the grid may be disconnected due to some abnormal conditions leading to islanding condition where the DG still continues to supply power the local loads. Similarly, there may arise fault disturbances due some type of faulty conditions between the phases and ground.

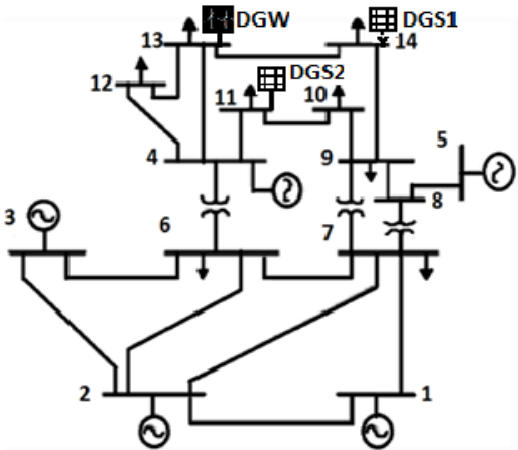

Figure 1. IEEE 14-bus power system with interconnection of DGs.

Fig. 1 shows the IEEE 14-bus power system with interconnection of DGs to some of its distribution buses. This model is developed in MATLAB/Simulink.

### 2.1 Wind energy conversion system (WECS)

Wind turbines are a non-linear, dynamic system that converts the wind kinetic energy into mechanical energy which is then processed to produce electrical energy by using wind generators. The characteristics of wind turbines are highly uncertain, non-linear and complex to design and study. Different models of wind turbines are proposed with many degrees of freedom (DOFs) to ensure the energy conversion. The mechanical power output from wind turbine in terms of wind speed is expressed as (Ackermann 2005).

$$P_{mech} = \frac{\rho}{2} A_{wind} C_{wind}(\eta, \beta) V_{wind}^3 \quad (1)$$

where $P_{mech}$ is the mechanical power output from the wind in watts; $\rho$ is the air density (kg/m3); $C_{wind}$ is the power coefficient; $\eta$ is the tip-speed ratio $(V_t/V_{wind})$, $V_t$ (m/s) is blade tip speed, and $V_{wind}$ (m/s) is wind speed; $A_{wind}$ is the area covered by rotor of wind turbine (m$^2$); and $\beta$ is the pitch angle (in degrees). The tip-speed ratio $\eta$ is given as

$$\eta = \frac{R\omega_r}{v_{wind}} \quad (2)$$

Where $\omega_r$ is the electrical speed (elec. rad/s). The doubly-fed induction generator (DFIG) model used as a part of the WECS (Tapia et al. 2003, Ray et al. 2010).

### 2.2 Photovoltaic (PV)

The solar PV module is formed by combination of cells represented as basic P-N junction diode. The PV cells convert the input solar radiation into electricity. PV cells are connected in a series and parallel combination to vary the output power ratings. The output voltage V and the load current I of PV system are expressed by:

$$I = I_L - I_0 \left[ \exp(\frac{V + IR_s}{\alpha}) - 1 \right] \quad (3)$$

Where $I_L$ is the PV current in amps; $I_0$ is the saturation current; $R_s$ is the series resistance in ohms and $\alpha$ is the thermal constant (Onara et al. 2008).

## 3 ISLANDING DETECTION METHODS

### 3.1 Wavelet transform (WT) as detection method

Wavelet is a tool which is applied to decompose a signal/function into different components and associate a frequency band to each of them. WT also changes the scales and frequencies of the analyzed signal and components. WT does not require a fixed basis function as in case of Fourier transform (FT), rather, different set of possible basis functions can be formulated to improve the performance.

In the present study, the voltage signals of the system is extracted and processed by Haar mother wavelet to decompose and detect the disturbances (Fernandez et al. 2002, Yadav et al. 2014). Of course, the voltage is filtered out by a series of low and high pass filters to obtain approximation (A) and detail (D) coefficients. The decomposition in of

approximation and detailed components are expressed as:

$$A_1(n) = \sum_k H(k-2n)\,C_0(k)\,;\; D_1(n) = \sum_k G(k-2n)\,C_0(k) \quad (4)$$

where, $H(n)$ is low-pass and $G(n)$ is the high-pass components of filter. For every decomposition scale, approximate and detail coefficients $(A_1(n)\,\&\,D_1(n))$ is determined for time-frequency analysis of the signal, $C_0(n)$.

### 3.2 STD and Energy as performance indices

Based on Parseval's theorem, it is known that energy of a signal $V(t)$ becomes same both in time as well as frequency domains and is given as

$$E_{signal} = \frac{1}{T}\int_0^T |V(t)|^2 \, dt = \sum_{n=0}^{K} |V[n]|^2 \quad (5)$$

where $T$ and $K$ are time period and signal length, respectively, and $V[n]$ is FT of the signal. Then, energy can be calculated by WT (Ackermann et al. 2005)

$$E_{signal} = \int |y(t)|^2 dt = \sum_{k=-\infty}^{\infty} |c(k)|^2 + \sum_{j=jo}^{\infty}\sum_{k=-\infty}^{\infty} |d_j(k)|^2 \quad (6)$$

The standard deviation and energy is calculated from the detail component (d1) of the voltage signal.

## 4 SIMULATED RESULTS

This section presents the simulation and analysis of the system and techniques for islanding and fault disturbance detection in power system using wavelet transform under different operating conditions. The study is being implemented in an IEEE 14-bus power system with DG penetration like wind and solar photovoltaic energy system. The models of the above power system are simulated in MATLAB/SIMULINK environment The parameters of various components used for simulation are given in the reference (Zeineldin et al. 2006, Jang et al. 2004, Dubey at al. 2011, Cheng et al. 2008).

### 4.1 Islanding detection

The islanding detection study using wavelet transform is presented in the 14-bus hybrid power system under different operating conditions. The voltage signal is captured from the bus-14 to which the solar DG1 (DGS1) system is connected along with the loads. The voltage signal is then being processed by wavelet transform to detect the islanding disturbance and the simulated results of the coefficients of Haar wavelet transform is shown in Figure 2. Of course, here it is observed that the signal is represented for 1000 sec, where as the approximate and detail coefficients are represented in 500 sec, because the 1000 sec window is divided equally. Therefore, the islanding event which starts at around 610 secs. is reflected to be detected by the coefficients at about half of 305 secs.

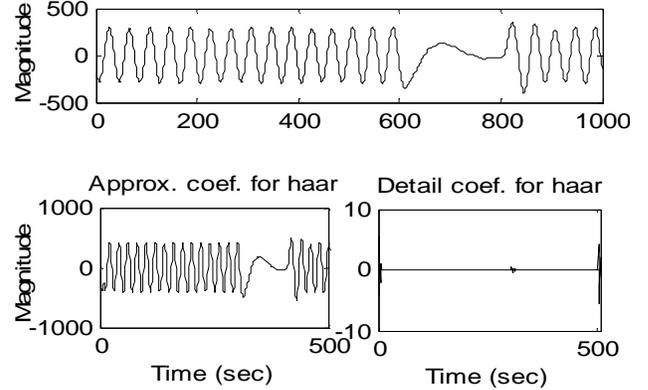

Figure 2. Islanding detection at bus-14 using wavelet transform.

### 4.2 Fault disturbance detection

This sub-section presents the study of detection of fault disturbances. Similar to the previous section, the voltage signal at bus-13, where the wind turbine is connected, is being measured and taken offline for processing. This signal is collected at the bus when there is an AG fault in the line connected between bus-13 and bus-14. Of course, both the voltages at the buses are affected. But, the voltage signal at bus-13 is taken into account for study in order to know the effect of fault as well as wind turbine on the voltage.

The voltage signal with its detection results by detail and approximate coefficient of Haar wavelet is presented in Figure 3. As seen from the figure that the fault occurs at about 60 secs. in the signal and the corresponding detection in the detail and approximate coefficients of Haar wavelet are reflected in figure. It is observed from the detail coefficient that the instant fault occurrence is clearly detected with an sudden increase and oscillation of magnitude and when the fault is cleared, the voltage and its coefficients become normal.

Similarly, the detection results for a AB fault in the connected line near bus-11 is shown in Figure 4. The voltage signal represents the variation which is also being detected by the wavelet coefficients. As soon as the fault is cleared, the signal as well as the coefficient comes to the normal operating conditions.

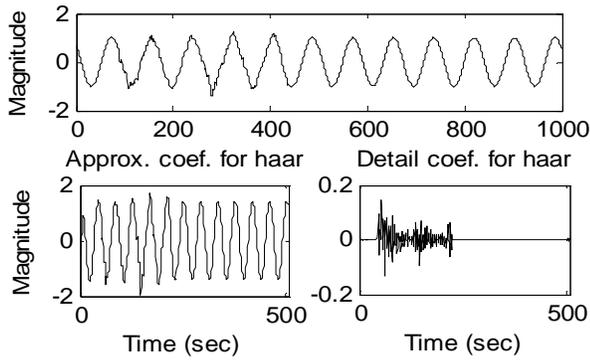

Figure 3. AG fault detection at bus-13 using wavelet transform.

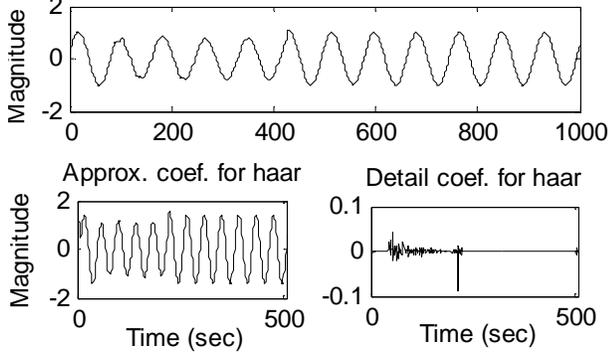

Figure 4. AB fault detection at bus-11 using wavelet transform.

### 4.3 *Energy and STD based detection of Islanding and Fault disturbances*

Table 1. STD and energy for islanding faults

(a) Haar wavelet transform

| Scenario | Fault disturbance | | Islanding | |
|---|---|---|---|---|
| | STD | Energy | STD | Energy |
| Bus-11 (AG) | $1.39 \times 10^{-5}$ | 0.6676 | $1.49 \times 10^{-5}$ | 1.4657 |
| Bus-12 (BG) | $2.44 \times 10^{-11}$ | 0.8462 | $1.69 \times 10^{-9}$ | 2.3645 |
| Bus-13 (CG) | $2.74 \times 10^{-5}$ | 1.5670 | $2.68 \times 10^{-4}$ | 4.6767 |
| Bus-11 (AB) | $1.39 \times 10^{-5}$ | 1.6874 | $2.20 \times 10^{-3}$ | 5.6786 |
| Bus-12 (BC) | $1.39 \times 10^{-5}$ | 0.9761 | $1.67 \times 10^{-5}$ | 4.4656 |
| Bus-13 (CA) | $2.30 \times 10^{-5}$ | 1.5731 | $2.33 \times 10^{-4}$ | 4.6763 |

(b) Daubechies4 (dB4) wavelet-transform

| Scenario | Fault disturbance | | Islanding | |
|---|---|---|---|---|
| | STD | Energy | STD | Energy |
| Bus-11 (AG) | $0.48 \times 10^{-5}$ | 0.0671 | $3.13 \times 10^{-5}$ | 1.1935 |
| Bus-12 (BG) | $1.83 \times 10^{-11}$ | 0.0892 | $2.37 \times 10^{-4}$ | 2.3843 |
| Bus-13 (CG) | $2.39 \times 10^{-5}$ | 0.2460 | $2.03 \times 10^{-5}$ | 3.0110 |
| Bus-11 (AB) | $1.23 \times 10^{-5}$ | 0.3834 | $2.35 \times 10^{-5}$ | 3.3921 |
| Bus-12 (BC) | $1.28 \times 10^{-5}$ | 0.2781 | $3.39 \times 10^{-5}$ | 2.9380 |
| Bus-13 (CA) | $2.06 \times 10^{-5}$ | 0.4830 | $2.29 \times 10^{-4}$ | 3.3921 |

(c) Comparison with different transforms

| Technique | Fault | | Islanding | |
|---|---|---|---|---|
| | STD | Energy | STD | Energy |
| FT | 0.453 | 0.967 | 1.120 | 1.894 |
| STFT | 0.634 | 1.034 | 1.673 | 2.171 |
| WT_dB1 | 0.842 | 1.495 | 1.784 | 2.742 |
| WT_Haar | 0.793 | 1.567 | 1.649 | 2.698 |
| WT_Coif | 0.830 | 1.675 | 1.801 | 2.822 |
| WT_Demey | 0.772 | 1.716 | 2.014 | 2.963 |

The detection of islanding as well as fault disturbances based on the value of performance indices (PIs) like standard deviation (STD) and energy, are presented in this sub-section. The indices are evaluated from the output coefficients of wavelet. These values are then compared with the selected threshold values to know whether they come greater/lower.

In case the index is higher than the threshold, islanding is identified, otherwise, if less, fault disturbance is detected. The computed energy and STD values are presented in Table 1. But, the challenge is there in selection of proper threshold value, which may be complex depending upon the type of network configuration as well as operating conditions. After finding the different values of PIs, as shown in Table, all the values are being compared with the threshold to identify the type of disturbance. A comparison between the different transforms is presented in (c) which suggest the increase in PI values in case of WTs, thereby increasing detection accuracy.

## 5 CONCLUSIONS

A study on detection of islanding as well as the fault disturbances in an IEEE 14-bus hybrid power system, in presence of DGs, is presented under different operating conditions. The disturbances are detected using Haar and dB4 wavelet transforms. Case studies are shown for identifying islanding and fault disturbances under various operating scenarios. Further, quantitative analysis using PIs in terms of STD and energy were also presented for disturbance detections. Both qualitative and quantitative analysis shows the effectiveness and accuracy of the proposed transform in detecting the islanding and faults.

## REFERENCES

Onara, O. C., M. Uzunoglua & M. S. Alam (2008). Modeling, control and simulation of an autonomous wind turbine/photovoltaic/fuel cell/ultra-capacitor hybrid power system. *J. Power Sources. 185 (2),* 1273-1283.

Kim, Seul-Ki, Jeon Jin-Hong, Cho Chang-Hee, Jong-Bo Ahn, & Kwon Sae-Hyuk (2008). Dynamic Modeling and Control of a Grid-Connected Hybrid Generation System With Versatile Power Transfer. *IEEE Trans. on Ind. Electr. 55 (4),* 1677-1688.

Ray, P. K., H. C. Dubey, S. R. Mohanty, Nand Kishor & K. Ganesh (2010). Power quality disturbance detection in grid-connected wind energy system using wavelet and S-transform. *IEEE International Conference on Power, Control and Embedded Systems (ICPCES), November 29-Dec. 1,* 1 – 4.

Fernandez ALO & NKI Ghonaim (2002). A novel approach using a FIRANN for fault detection and direction estimation for high voltage transmission lines. *IEEE Trans Power Deliv. 17,* 894–901.

Yadav A & A. Swetapadma (2014). Improved first zone reach setting of artificial neural network-based directional relay for protection of double circuit transmission lines. *IET Gen Transm Distrib 8(3),* 373–88.

Zeineldin H. H, Ehab F. El-Saadany & MMA. Salama (2006). Impact of DG interface control on islanding detection and non-detection zones. *IEEE Trans Power Delivery 21(3)*, 1515–1523.

Jang S & K. Kim (2004). An islanding detection method for distributed generation algorithm using voltage unbalance and total harmonic distortion of current. *IEEE Trans Power Delivery 19(2)*, 745–752.

Dubey H. C., S. R. Mohanty, Nand Kishor & P. K. Ray (2011). Fault Detection in a Series Compensated Transmission Line using Discrete Wavelet Transform and Independent Component Analysis: a Comparative Study. *5th International Power Engineering and Optimization Conference (PEOCO), Shah Alam, Selangor, Malaysia*.

Cheng-Tao Hsieh, Jeu-Min Lin & Shyh-Jier Huang (2008). Enhancement of islanding-detection of distributed generation systems via wavelet transform-based approaches. J. *Electr. Power and Energy Syst., 30,* 575–580.

Ackermann T. (2005) *Wind power in power systems* Chichester, Wiley.

Tapia, A., G. Tapia, J. X. Ostolaza & J. R. Saenz ( 2003). Modeling control of a wind turbine driven doubly fed induction generator. *IEEE Trans. Energy Convers. 18 (2),* 194–204.

Ray P. K., S. R. Mohanty & Nand Kishor (2010). Coherency determination in grid-connected distributed generation based hybrid system under islanding scenarios. *IEEE International Conference on Power and Energy (PECON), Kuala Lumpur, Malaysia, Nov 29-Dec. 1,* 10, 85 – 88.